\documentclass[pra,showpacs,twocolumn,superscriptaddress]{revtex4-1}
\usepackage{amssymb}
\usepackage{amsmath}
\usepackage{graphicx}
\usepackage{epsfig}

\begin{document}

\title{EPR pairing dynamics in Hubbard model with resonant $U$}
\author{X. Z. Zhang}
\affiliation{School of Physics, Nankai University, Tianjin 300071, China}
\affiliation{College of Physics and Materials Science, Tianjin Normal University, Tianjin
300387, China}
\author{Z. Song}
\email{songtc@nankai.edu.cn}
\affiliation{School of Physics, Nankai University, Tianjin 300071, China}

\begin{abstract}
We study the dynamics of the collision between two fermions in Hubbard model
with on-site interaction strength $U$. The exact solution shows that the
scattering matrix for two-wavepacket collision is separable into two
independent parts, operating on spatial and spin degrees of freedom,
respectively. The S-matrix for spin configuration is equivalent to that of
Heisenberg-type pulsed interaction with the strength depending on $U$ and
relative group velocity $\upsilon _{r}$. This can be applied to create
distant EPR pair, through a collision process for two fermions with opposite
spins in the case of $\left\vert \upsilon _{r}/U\right\vert =1$,\ without
the need for temporal control and measurement process. Multiple collision
process for many particles is also discussed.
\end{abstract}

\maketitle


\section{Introduction}

\label{sec_intro}Pairing is the origin of many fascinating phenomena in
nature, ranging from superconductivity to quantum teleportation. Owing to
the rapid advance of experimental techniques, it has been possible both to
produce Cooper pairs of fermionic atoms and to observe the crossover between
a Bose-Einstein condensate and a Bardeen-Cooper-Schrieffer superfluid \cite%
{Regal,Zwierlein,Bourdel}. The dynamic process of pair formation is of
interest in both condensed matter physics and quantum information science.
On one hand, the collective behavior of pairs gives rise to macroscopic
properties in many-body physics. On the other hand, a single entangled pair
is a promising quantum information resource for future quantum computation.

In recent years, the controlled setting of ultracold fermionic atoms in
optical lattices is regarded as a promising route to enabled quantitative
experimental tests of theories of strongly interacting fermions \cite%
{ReviewBloch,Zwerger,ZwierleinOX,NPBloch}.\ In particular, fermions trapped
in optical lattices can directly simulate the physics of electrons in a
crystalline solid, shedding light on novel physical phenomena in materials
with strong electron correlations \cite%
{ReviewBloch,Lewenstein,EsslingerAnnual}. A major effort is devoted to
simulate the Fermi-Hubbard model by using ultracold neutral atoms \cite%
{Byrnes1,Byrnes2,Murmann}. This approach offers experimental access to a
clean and highly flexible Fermi-Hubbard model with a unique set of
observables \cite{Esslinger} and therefore, motivate a large number of works
on Mott insulator phase \cite{Jordens,Schneider} and transport properties
\cite{Strohmaier,Hackermller}, stimulating further theoretical and
experimental investigations on the dynamics of strongly interacting
particles for the Fermi Hubbard model.

In this paper, we study the dynamics of the collision between two fermions
with various spin configurations. The particle-particle interaction is
described by Hubbard model, which operates spatial and spin degrees of
freedom in a mixed manner. Based on the Bethe ansatz solution, the time
evolution of two fermonic wave packets with identical size is analytically
obtained. We find that the scattering matrix of the collision is separable
into two independent parts, operating on spatial and spin degrees of
freedom, respectively. The scattered two particles exhibit dual features.
The spatial part behaves as classical particles, swapping the momenta, while
the spin part obeys the isotropic Heisenberg-type exchange coupling. The
coupling strength depends on the Hubbard on-site interaction and relative
group velocity of two wavepackets.\ This finding can be applied to create
distant EPR pair, through a collision process for two fermions with opposite
spins\ without the need for temporal control and measurement process.
Multiple collision process for many particles is also discussed.

The paper is organized as follow. In Sec. \ref{sec_model}, we present the
model Hamiltonian and analyze the symmetries. In Sec. \ref{sec_two-particles}%
, we investigate some exact results obtained by Bethe ansatz concerning the
two-particle problem. In Sec. \ref{sec_dynamics}, we explore the dynamics of
wavepacket collision.\ Sec. \ref{sec_Heisenberg} is devoted to construct the
scattering matrix for\ the collision process and the corresponding
equivalent Hamiltonian. In Sec. \ref{sec_multiple}, we apply two-particle
S-matrix to the case of multi-fermion collision. Finally, we give a summary
and discussion in Sec. \ref{sec_summary}.

\section{Model Hamiltonians and symmetries}

\label{sec_model}A one-dimensional Hubbard Hamiltonian on an $N$-site ring
reads
\begin{equation}
H=-\kappa \sum_{i=1,\sigma =\uparrow \downarrow }^{N}\left( c_{i,\sigma
}^{\dagger }c_{i+1,\sigma }+\text{H.c.}\right) +U\sum_{i}n_{i\uparrow
}n_{i\downarrow },  \label{F_Hub}
\end{equation}%
where $c_{i,\sigma }^{\dagger }$ is the creation operator of the fermion at
the site $i$ with spin $\sigma =\uparrow ,\downarrow $ and $U$ is the
on-site interaction. The tunneling strength and the on-site interaction
between bosons are denoted by $\kappa $ and $U$. For the sake of clarity and
simplicity, we only consider odd-site system with\ $N=2N_{0}+1$, and
periodic boundary condition\ $c_{i,\sigma }=c_{i+N,\sigma }$.

We analyze three symmetries of the Hamiltonian as following, which is
critical for achieving a two-particle solution. The first is particle-number
conservation $\left[ N_{\sigma },H\right] =0$, where $N_{\sigma
}=\sum_{i}c_{i,\sigma }^{\dagger }c_{i,\sigma }$, which ensures that one can
solve the eigen problem in the invariant subspace with fixed $N_{\sigma }$,
no matter $U$ is real or complex. The second is the translational symmetry, $%
\left[ T_{1},H\right] =0$, where $T_{1}$ is the shift operator defined as
\begin{eqnarray}
T_{1}^{-1}c_{i,\sigma }^{\dagger }T_{1} &=&c_{i+1,\sigma }^{\dagger },
\notag \\
\text{or }T_{1}^{-1}c_{k,\sigma }^{\dagger }T_{1} &=&e^{-ik}c_{k,\sigma
}^{\dagger },
\end{eqnarray}%
with
\begin{eqnarray}
c_{k,\sigma }^{\dagger } &=&\frac{1}{\sqrt{N}}\sum_{j}e^{ikj}c_{j,\sigma
}^{\dagger },  \notag \\
k &=&2n\pi /N\,n\in \left[ 1,N\right] .
\end{eqnarray}%
This allows invariant subspace spanned by the eigenvector of operator $T_{1}$%
. Based on this fact, one can reduce the two-particle problem to a
single-particle problem.\textbf{\ }The final is the SU(2) symmetry\textbf{,}
$\left[ S^{\pm ,z},H\right] =0$ and $\left[ S^{2},H\right] =0$, where the
spin operators are defined as%
\begin{eqnarray}
S^{+} &=&\left( S^{-}\right) ^{\dag }=\sum_{i}c_{i,\uparrow }^{\dagger
}c_{i,\downarrow }, \\
S^{z} &=&\frac{1}{2}\sum_{i}\left( c_{i,\uparrow }^{\dagger }c_{i,\uparrow
}-c_{i,\downarrow }^{\dagger }c_{i,\downarrow }\right) ,
\end{eqnarray}%
which satisfy the relation $\left[ S^{+},S^{-}\right] =2S^{z}$.

Now based on the above analysis, we construct the basis of the two-fermion
invariant subspace as following
\begin{eqnarray}
&&\left\vert \phi _{0}^{-}\left( K\right) \right\rangle =\frac{1}{\sqrt{N}}%
\sum_{j}e^{iKj}c_{j,\uparrow }^{\dagger }c_{j,\downarrow }^{\dagger
}\left\vert \text{vac}\right\rangle ,  \label{bases1} \\
&&\left\vert \phi _{r}^{\pm }\left( K\right) \right\rangle =\frac{1}{\sqrt{2N%
}}e^{iKr/2}\sum_{j}e^{iKj}  \notag \\
&&\times \left( c_{j,\uparrow }^{\dagger }c_{j+r,\downarrow }^{\dagger }\pm
c_{j,\downarrow }^{\dagger }c_{j+r,\uparrow }^{\dagger }\right) \left\vert
\text{vac}\right\rangle \text{, }\left( r>1\right) ,  \notag
\end{eqnarray}%
and%
\begin{equation}
\frac{S^{\pm }}{\sqrt{2}}\left\vert \phi _{r}^{+}\left( K\right)
\right\rangle =\frac{1}{\sqrt{N}}e^{iKr/2}\sum_{j}e^{iKj}c_{j,\pm \uparrow
}^{\dagger }c_{j+r,\pm \uparrow }^{\dagger }\text{, }\left( r>1\right) ,
\label{bases2}
\end{equation}%
where $K$\ is the momentum vector, indexing the subspace. These bases are
eigenvectors of the operators $N_{\sigma }$, $T_{1}$,\ $S^{2}$\ and $S^{z}$.
Straightforward algebra yields%
\begin{eqnarray}
N_{\sigma }\left\vert \phi _{0}^{-}\left( K\right) \right\rangle
&=&\left\vert \phi _{0}^{-}\left( K\right) \right\rangle ,\text{ } \\
N_{\sigma }\left\vert \phi _{r}^{\pm }\left( K\right) \right\rangle
&=&\left\vert \phi _{r}^{\pm }\left( K\right) \right\rangle , \\
N_{\uparrow }\frac{S^{\pm }}{\sqrt{2}}\left\vert \phi _{r}^{+}\left(
K\right) \right\rangle &=&\left( 1\pm 1\right) \frac{S^{\pm }}{\sqrt{2}}%
\left\vert \phi _{r}^{+}\left( K\right) \right\rangle ,
\end{eqnarray}%
and%
\begin{eqnarray}
T_{1}\left\vert \phi _{0}^{-}\left( K\right) \right\rangle
&=&e^{-iKj}\left\vert \phi _{0}^{-}\left( K\right) \right\rangle ,\text{ } \\
T_{1}\left\vert \phi _{r}^{\pm }\left( K\right) \right\rangle
&=&e^{-iKj}\left\vert \phi _{r}^{\pm }\left( K\right) \right\rangle , \\
T_{1}S^{\pm }\left\vert \phi _{r}^{+}\left( K\right) \right\rangle
&=&e^{-iKj}S^{\pm }\left\vert \phi _{r}^{+}\left( K\right) \right\rangle ,
\end{eqnarray}%
while%
\begin{eqnarray}
S^{2}\left\vert \phi _{0}^{-}\left( K\right) \right\rangle
&=&S^{2}\left\vert \phi _{r}^{-}\left( K\right) \right\rangle =0,\text{ } \\
S^{2}\left\vert \phi _{r}^{+}\left( K\right) \right\rangle &=&2\left\vert
\phi _{r}^{+}\left( K\right) \right\rangle , \\
S^{2}S^{\pm }\left\vert \phi _{r}^{+}\left( K\right) \right\rangle
&=&2S^{\pm }\left\vert \phi _{r}^{+}\left( K\right) \right\rangle ,
\end{eqnarray}%
and%
\begin{eqnarray}
S^{z}\left\vert \phi _{0}^{-}\left( K\right) \right\rangle
&=&S^{z}\left\vert \phi _{r}^{\pm }\left( K\right) \right\rangle =0, \\
S^{z}S^{\pm }\left\vert \phi _{r}^{+}\left( K\right) \right\rangle &=&\pm
S^{\pm }\left\vert \phi _{r}^{+}\left( K\right) \right\rangle .
\end{eqnarray}%
Then there are four invariant subspaces with $\left( S,S^{z}\right) =\left(
0,0\right) $, $\left( 1,0\right) $, and $\left( 1,\pm 1\right) $\ involved.

\begin{figure}[tbp]
\includegraphics[ bb=31 400 544 757, width=0.45\textwidth, clip]{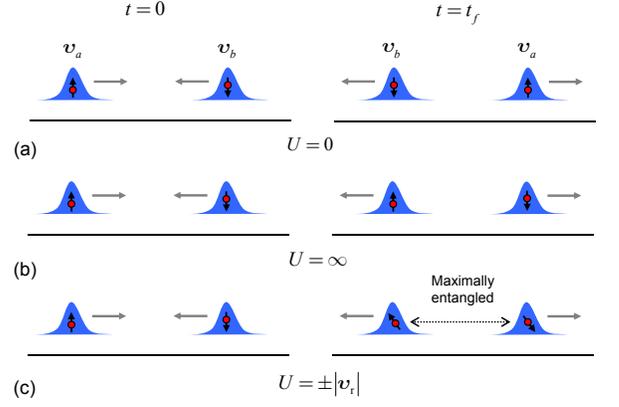}
\caption{(Color online) Schematic illustration of the collision process of
two separated fermionic wavepackets with opposite spin orientations for
three typical values of $U$. In all cases, the collisions result in momentum
swap, but different spin configurations: (a) $U=0$, swap; (b) $U=\infty $,
unchange; (c) $U=\pm \left\vert \protect\upsilon _{r}\right\vert $, maximal
entanglement. } \label{fig1}
\end{figure}


\section{Two-particle solutions}

\label{sec_two-particles}Now we look at the two-particle solution in each
invariant subspace. We only focus on the solutions in subspaces $\left(
0,0\right) $ and $\left( 1,0\right) $, since the one in subspace $\left(
1,\pm 1\right) $ can be obtained directly from that in subspace $\left(
1,0\right) $ by operator $S^{\pm }$. A two-particle state can be written as%
\begin{equation}
\left\vert \varphi _{K}^{\pm }\right\rangle =\sum_{r}f_{K,k}^{\pm }\left(
r\right) \left\vert \phi _{r}^{\pm }\left( K\right) \right\rangle \text{, }%
\left( f_{K}^{+}\left( 0\right) =f_{K,k}^{-}\left( -1\right) =0\right) ,
\end{equation}%
where the wave function $f_{K,k}^{\pm }\left( r\right) $\ satisfies the Schr%
\"{o}dinger equations%
\begin{eqnarray}
Q_{r}^{K}f_{K,k}^{+}\left( r+1\right) +Q_{r-1}^{K}f_{K,k}^{+}\left(
r-1\right) + &&  \notag \\
\lbrack \left( -1\right) ^{n}Q_{r}^{K}\delta _{r,N_{0}}-\varepsilon
_{K}]f_{K,k}^{+}\left( r\right) =0, &&  \label{S+}
\end{eqnarray}%
and%
\begin{eqnarray}
Q_{r}^{K}f_{K,k}^{-}\left( r+1\right) +Q_{r-1}^{K}f_{K,k}^{-}\left(
r-1\right) + &&  \notag \\
\lbrack U\delta _{r,0}+\left( -1\right) ^{n}Q_{r}^{K}\delta
_{r,N_{0}}-\varepsilon _{K}]f_{K,k}^{-}\left( r\right) =0, &&  \label{S-}
\end{eqnarray}%
with the eigen energy $\varepsilon _{K}$\ in the invariant subspace indexed
by $K$. Here factor $Q_{r}^{K}=-2\sqrt{2}\kappa \cos \left( K/2\right) $ for
$r=0$ and $-2\kappa \cos \left( K/2\right) $ for $r\neq 0$, respectively. As
pointed in Ref. \cite{L. Jin} in previous works, the eigen problem of
two-particle matrix can be reduced to the that of single particle. We see
that the solution of (\ref{S-}) is equivalent to that of the single-particle
$N_{0}+1$-site tight-binding chain system with nearest-neighbour (NN)
hopping amplitude $Q_{j}^{K}$, on-site potentials $U$ and $\left( -1\right)
^{n+1}2\kappa \cos \left( K/2\right) $ at $0$th and $N_{0}$th sites,
respectively. The solution of (\ref{S+}) corresponds to the same chain with
infinite $U$. In this work, we only concern the scattering solution by the $%
0 $th end. In this sense, $f_{K}^{-}$ can be obtained from the equivalent
Hamiltonian%
\begin{equation}
H_{\text{eq}}^{K}=U\left\vert 0\right\rangle \left\langle 0\right\vert
+\sum_{i=1}^{\infty }\left( Q_{i}^{K}\left\vert i\right\rangle \left\langle
i+1\right\vert +\text{H.c.}\right) .  \label{H_eq}
\end{equation}%
%
%
%
%
%
%
%
%

\begin{figure}[tbp]
\includegraphics[ bb=33 165 542 602, width=0.4\textwidth, clip]{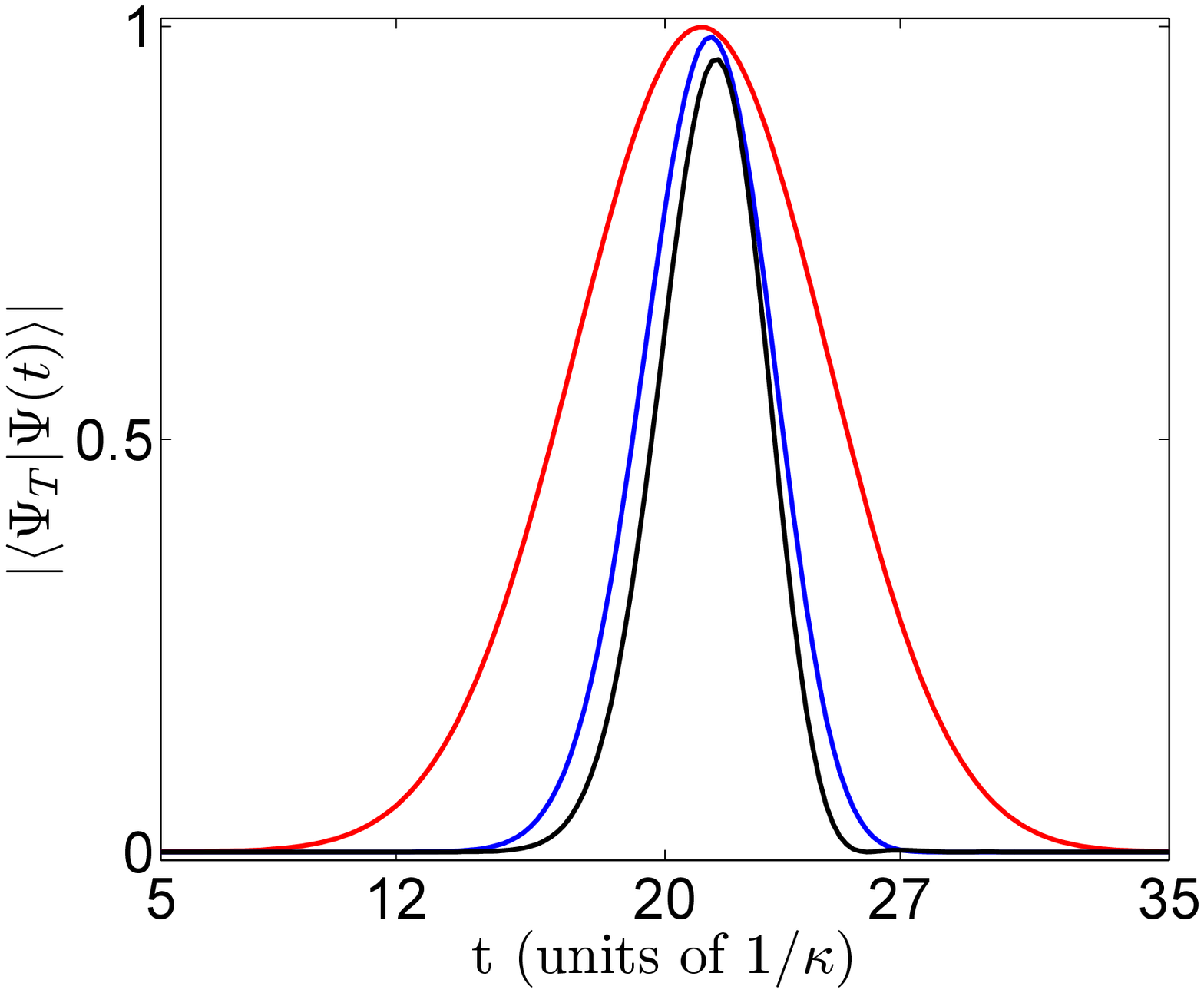}
\caption{(Color online) Plots of the fidelity $\left\vert \left\langle \Psi
_{T}\right\vert \Psi \left( t\right) \rangle \right\vert $ with the
parameters $N_{a}=20,N_{b}=62$, $k_{a}=-k_{b}=\protect\pi /2$, in the system
Eq. (\protect\ref{F_Hub}) with $N=81$ and $U=\protect\upsilon _{r}$. The
red, blue and black lines represent the plots of fidelity $\left\vert
\left\langle \Psi _{T}\right\vert \Psi \left( t\right) \rangle \right\vert $
in the condition of $\protect\alpha =0.13$, $0.26$, and $0.33$,
respectively. It shows that the fidelity is close $1$, as $\protect\alpha $
approaches to $0$, which accords with the theoretical analysis in the text.} %
\label{fig2}
\end{figure}

Based on the Bethe ansatz technique, the scattering solution can be
expressed as%
\begin{equation}
f_{K,k}^{-}\left( j\right) =e^{-ikj}+R_{K,k}e^{ikj},  \label{fkj}
\end{equation}%
with eigen energy $\varepsilon _{K}\left( k\right) =-4\kappa \cos \left(
K/2\right) \cos k$, $k\in \left[ 0,\pi \right] $. Here the reflection
amplitude%
\begin{equation}
R_{K,k}=\frac{i\lambda _{K,k}+U}{i\lambda _{K,k}-U}=e^{i\Delta },  \label{R}
\end{equation}%
where%
\begin{eqnarray}
\lambda _{K,k} &=&4\kappa \cos \left( K/2\right) \sin k,  \label{U_c} \\
\Delta  &=&2\tan ^{-1}\left( -\frac{U}{\lambda _{K,k}}\right) .
\end{eqnarray}%
And $f_{K,k}^{+}$\ can be obtained from $f_{K,k}^{-}$\ by taking\textbf{\ }$%
U=\infty $\textbf{.} We note that $R_{K,k}\left( -U\right) =R_{K,k}^{\ast
}=R_{K,-k}$, which reveals a dynamic symmetry of the Hubbard model with
respect to the sign of $U$\textbf{.}

\section{Dynamics of wavepacket collision}

\label{sec_dynamics}In this section, we investigate the dynamics of
two-wavepackets collision based on the above formalism. We begin with our
investigation from the time evolution of an initial state%
\begin{equation}
\left\vert \Phi \right\rangle =\left\vert \Phi _{a,\sigma }\right\rangle
\left\vert \Phi _{b,\sigma ^{\prime }}\right\rangle ,  \label{Phi}
\end{equation}%
which represents two separable fermions $a$ and $b$, with spin $\sigma $\
and $\sigma ^{\prime }$, respectively. Here%
\begin{equation}
\left\vert \Phi _{\gamma ,\sigma }\right\rangle =\frac{1}{\sqrt{\Omega }}%
\sum_{j}e^{-\alpha ^{2}\left( j-N_{\gamma }\right) ^{2}}e^{ik_{\gamma
}j}c_{j,\sigma }^{\dagger }\left\vert \text{Vac}\right\rangle ,
\end{equation}%
with $\gamma =a$, $b$ and $N_{a}-N_{b}\gg 1/\alpha $, is a wavepacket with a
width $2\sqrt{\ln 2}/\alpha $, a central position $N_{\gamma }$ and a group
velocity $\upsilon _{\gamma }=-2\kappa \sin k_{\gamma }$. We focus on the
case $\left( \sigma ,\sigma ^{\prime }\right) =\left( \uparrow ,\downarrow
\right) $. The obtained result can be extended to other cases. In order to
calculate the time evolution of state $\left\vert \Phi \right\rangle $, two
steps are necessary. At first, the projection of $\left\vert \Phi
\right\rangle $ on the basis sets $\left\{ \left\vert \phi _{r}^{+}\left(
K\right) \right\rangle \right\} $ and $\left\{ \left\vert \phi
_{r}^{-}\left( K\right) \right\rangle \right\} $\ can be given by the
decomposition%
\begin{eqnarray}
2\left\vert \Phi _{a,\uparrow }\right\rangle \left\vert \Phi _{b,\downarrow
}\right\rangle &=&\left( \left\vert \Phi _{a,\uparrow }\right\rangle
\left\vert \Phi _{b,\downarrow }\right\rangle +\left\vert \Phi
_{a,\downarrow }\right\rangle \left\vert \Phi _{b,\uparrow }\right\rangle
\right)  \notag \\
&&+\left( \left\vert \Phi _{a,\uparrow }\right\rangle \left\vert \Phi
_{b,\downarrow }\right\rangle -\left\vert \Phi _{a,\downarrow }\right\rangle
\left\vert \Phi _{b,\uparrow }\right\rangle \right) .
\end{eqnarray}%
Secondly, introducing the transformation%
\begin{eqnarray}
N_{c} &=&\frac{1}{2}\left( N_{a}+N_{b}\right) ,r_{c}=N_{b}-N_{a}, \\
k_{c} &=&\frac{1}{2}\left( k_{a}+k_{b}\right) ,q_{c}=k_{b}-k_{a}, \\
l &=&j+r,
\end{eqnarray}%
and using the identities
\begin{eqnarray}
&&2\left[ \left( j-N_{a}\right) ^{2}+\left( l-N_{b}\right) ^{2}\right]
\notag \\
&=&\left[ \left( j+l\right) -\left( N_{a}+N_{b}\right) \right] ^{2}+\left[
\left( l-j\right) -\left( N_{b}-N_{a}\right) \right] ^{2}, \\
&&2\left( k_{a}j+k_{b}l\right)  \notag \\
&=&\left( k_{a}+k_{b}\right) \left( j+l\right) +\left( k_{b}-k_{a}\right)
\left( l-j\right) ,
\end{eqnarray}%
we have%
\begin{eqnarray}
&&\frac{1}{\sqrt{2}}\left( \left\vert \Phi _{a,\uparrow }\right\rangle
\left\vert \Phi _{b,\downarrow }\right\rangle \pm \left\vert \Phi
_{a,\downarrow }\right\rangle \left\vert \Phi _{b,\uparrow }\right\rangle
\right)  \notag \\
&=&\frac{1}{\sqrt{\Omega _{1}}}\sum_{K}e^{-\left( K-2k_{c}\right)
^{2}/4\alpha ^{2}}  \notag \\
&&\times e^{-iN_{c}\left( K-2k_{c}\right) }\left\vert \psi _{K}^{\pm }\left(
r_{c},q_{c}\right) \right\rangle ,
\end{eqnarray}%
with%
\begin{equation}
\left\vert \psi _{K}^{\pm }\right\rangle =\frac{1}{\sqrt{\Omega _{2}}}%
\sum_{r}e^{-\alpha ^{2}\left( r-r_{c}\right) ^{2}/2}e^{iq_{c}r/2}\left\vert
\phi _{r}^{\pm }\left( K\right) \right\rangle ,
\end{equation}%
where $\Omega _{1,2}$ is the normalized factor.

We note that the component of state $\left\vert \Phi \right\rangle $\ on
each invariant subspace indexed by $K$ represents an incident wavepacket
along the chain described by\ $H_{\text{eff}}^{K}$. This wavepacket has
width $2\sqrt{\ln 2}/\alpha $, central position $r_{c}=N_{b}-N_{a}$\ and
group velocity $\upsilon =-4\kappa \cos \left( K/2\right) \sin \left(
q_{c}/2\right) $.\ Accordingly, the time evolution of state $\left\vert \Phi
\right\rangle $\ can be derived by the dynamics of each sub wavepacket in
each chain $H_{\text{eff}}^{K}$, which eventually can be obtained from Eq. (%
\ref{fkj}).\ According to the solution, the evolved state of $\left\vert
\psi _{K}^{\pm }\left( r_{c},q_{c}\right) \right\rangle $ can be expressed
approximately in the form $e^{i\beta \left( r_{c}^{\prime }\right)
}R_{2k_{c},q_{c}/2}\left\vert \psi _{K}^{\pm }\left( r_{c}^{\prime
},-q_{c}\right) \right\rangle $, which represents a reflected wavepacket.
Here $\beta \left( r_{c}^{\prime }\right) $ is an overall phase, as a
function of $r_{c}^{\prime }$,\ the position of the reflected wavepacket,
being independent of $U$. In addition, it is easy to check out that, in the
case with\ $\alpha \ll 1$, the initial state distribute mainly in the
invariant subspace\ $K=2k_{c}$, where the wavepacket moves with the group
velocity\ $\upsilon _{\mathrm{r}}=-4\kappa \cos \left( k_{c}\right) \sin
\left( q_{c}/2\right) $ $=\upsilon _{b}-\upsilon _{a}$. Then the state after
collision has the approximate form
\begin{eqnarray}
\left\vert \Phi \left( \infty \right) \right\rangle &=&\frac{%
1-R_{2k_{c},q_{c}/2}}{\sqrt{\Omega }}(\sum_{j}e^{-\alpha ^{2}\left(
j-N_{a}^{\prime }\right) ^{2}}e^{ik_{b}j}c_{j,\uparrow }^{\dagger
}\left\vert \text{Vac}\right\rangle )  \notag \\
&&(\sum_{l}e^{-\alpha ^{2}\left( l-N_{b}^{\prime }\right)
^{2}}e^{ik_{a}l}c_{l,\downarrow }^{\dagger }\left\vert \text{Vac}%
\right\rangle )  \notag \\
&&+\frac{1+R_{2k_{c},q_{c}/2}}{\sqrt{\Omega }}(\sum_{j}e^{-\alpha ^{2}\left(
j-N_{a}^{\prime }\right) ^{2}}e^{ik_{b}j}c_{j,\downarrow }^{\dagger
}\left\vert \text{Vac}\right\rangle )  \notag \\
&&(\sum_{l}e^{-\alpha ^{2}\left( l-N_{b}^{\prime }\right)
^{2}}e^{ik_{a}l}c_{l,\uparrow }^{\dagger }\left\vert \text{Vac}\right\rangle
),
\end{eqnarray}%
which also represents two separable wavepackets at $N_{a}^{\prime }$\ and $%
N_{b}^{\prime }$\ respectively.\ Here $\Omega $\ is the normalized factor
and an overall phase is neglected\textbf{.}\ We would like to point that the
obtained conclusion is based on the fact that the shapes of two wavepackets $%
\left\vert \Phi _{a,\sigma }\right\rangle $ and $\left\vert \Phi _{b,\sigma
^{\prime }}\right\rangle $\ are identical.

\section{Equivalent Heisenberg coupling}

\label{sec_Heisenberg}Now we try to express the two-fermion collision in a
more compact form. We will employ an S-matrix to relate the asymptotic spin
states of the incoming to outcoming particles. We denote an incident
single-particle wavepacket as the form of $\left\vert \lambda ,p,\sigma
\right\rangle $, where $\lambda =\mathrm{L},$ $\mathrm{R}$\ indicates the
particle in the left and right of the collision zone, $p$ the momentum, and $%
\sigma =\uparrow ,\downarrow $\ the spin degree of freedom. In this context,
we give the asymptotic expression for the collision process as%
\begin{equation}
\left\vert \mathrm{L},p,\sigma _{\mathrm{L}}\right\rangle \left\vert \mathrm{%
R},q,\sigma _{\mathrm{R}}\right\rangle \longmapsto \mathcal{S}\left\vert
\mathrm{L},q,\sigma _{\mathrm{L}}\right\rangle \left\vert \mathrm{R}%
,p,\sigma _{\mathrm{R}}\right\rangle ,  \label{lambda_d}
\end{equation}%
where the S-matrix%
\begin{equation}
\mathcal{S}=e^{-i\left( \theta -\pi \right) (\overrightarrow{s}_{\mathrm{L}%
}\cdot \overrightarrow{s}_{\mathrm{R}}-1/4)},  \label{s-matrix}
\end{equation}%
governs the spin part of the wave function. Here $\overrightarrow{s}_{%
\mathrm{L,R}}$ denotes spin operator for the spins of particles at left or
right, $\theta =2\tan ^{-1}\left[ U/\left( \upsilon _{\text{\textrm{R}}%
}-\upsilon _{\text{\textrm{L}}}\right) \right] $\textbf{, }where $\upsilon _{%
\text{\textrm{L}}}$\ and $\upsilon _{\text{\textrm{R}}}$\ represent the
group velocity of the left and right wavepacket, respectively.\textbf{\ }%
Together with the scattering matrix $\mathcal{M}$ for spatial degree of
freedom%
\begin{equation}
\mathcal{M}\left\vert \mathrm{L},p,\sigma _{\mathrm{L}}\right\rangle
\left\vert \mathrm{R},q,\sigma _{\mathrm{R}}\right\rangle =\left\vert
\mathrm{L},q,\sigma _{\mathrm{L}}\right\rangle \left\vert \mathrm{R}%
,p,\sigma _{\mathrm{R}}\right\rangle ,
\end{equation}%
we have a compact expression%
\begin{equation}
\left\vert \mathrm{f}\right\rangle =\mathcal{MS}\left\vert \mathrm{i}%
\right\rangle ,  \label{MS}
\end{equation}%
to connect the initial and final states. In general the total scattering
matrix has the form \textrm{exp}$\left[ -i\int_{-\infty }^{\infty }H\mathrm{d%
}t\right] $, which is not separable into spatial and spin parts. Then Eq. (%
\ref{MS}) is only available for some specific initial states, e.g.,
spatially separable two-particle wavepackets with identical size. This may
lead to some interesting phenomena.

It is interesting to note that the scattering matrix for spin is equivalent
to the propagator%
\begin{equation}
\mathcal{S}=\mathcal{T}\mathrm{exp}\left[ -i\int_{-\infty }^{\infty }h\left(
t\right) \mathrm{d}t\right] ,
\end{equation}%
for a pulsed Hensenberg model with Hamiltonian%
\begin{equation}
h\left( t\right) =J\left( t\right) (\overrightarrow{s}_{\mathrm{L}}\cdot
\overrightarrow{s}_{\mathrm{R}}-1/4),
\end{equation}%
with $\int J\left( t\right) dt=\theta -\pi $. Here $\mathcal{T}$ is
time-ordered operator, which can be ignored since only the coupling strength
$J\left( t\right) $\ is time dependent. This observation accords with the
fact that, in the large positive $U$ case, the Hubbard model scales on the $%
t-J$ model \cite{Spalek,Auerbach}, which also includes the NN interaction
term of isotropic Heisenberg type.

This also indicates that the effect of collision on two spins is equivalent
to that of time evolution operation under the Hamiltonian $\overrightarrow{s}%
_{\mathrm{L}}\cdot \overrightarrow{s}_{\mathrm{R}}$\ at an appropriate
instant. In this sense, the collision process can be utilized to implement
two-qubit gate. For two coupled-qubit system, the time evolution operator is
simply given by
\begin{equation}
\mathcal{U}\left( t\right) =\exp \left( -i\overrightarrow{s}_{\mathrm{L}%
}\cdot \overrightarrow{s}_{\mathrm{R}}t\right) ,
\end{equation}%
yielding%
\begin{equation}
\mathcal{U}\left( t\right) \left\vert \uparrow \right\rangle _{\mathrm{L}%
}\left\vert \downarrow \right\rangle _{\mathrm{R}}=e^{it/4}(\cos \frac{t}{2}%
\left\vert \uparrow \right\rangle _{\mathrm{L}}\left\vert \downarrow
\right\rangle _{\mathrm{R}}-i\sin \frac{t}{2}\left\vert \downarrow
\right\rangle _{\mathrm{L}}\left\vert \uparrow \right\rangle _{\mathrm{R}}),
\end{equation}%
where $\left\vert \sigma =\uparrow ,\downarrow \right\rangle _{\mathrm{L,R}}$
denotes qubit state. We can see that at instants $t=\pi /2$ and $\pi $, the
evolved states become%
\begin{eqnarray}
\mathcal{U}\left( \pi /2\right) \left\vert \uparrow \right\rangle _{\mathrm{L%
}}\left\vert \downarrow \right\rangle _{\mathrm{R}} &=&\frac{e^{i\pi /8}}{%
\sqrt{2}}\left( \left\vert \uparrow \right\rangle _{\mathrm{L}}\left\vert
\downarrow \right\rangle _{\mathrm{R}}-i\left\vert \downarrow \right\rangle
_{\mathrm{L}}\left\vert \uparrow \right\rangle _{\mathrm{R}}\right) , \\
\mathcal{U}\left( \pi \right) \left\vert \uparrow \right\rangle _{\mathrm{L}%
}\left\vert \downarrow \right\rangle _{\mathrm{R}} &=&e^{-i\pi /4}\left\vert
\downarrow \right\rangle _{\mathrm{L}}\left\vert \uparrow \right\rangle _{%
\mathrm{R}},
\end{eqnarray}%
\ which indicates that $U\left( \pi /2\right) $\ and $U\left( \pi \right) $\
are entangling and\ swap operators, respectively. In practice, such
protocols require exact time control of the operation.

Comparing operator $U\left( t\right) $\ and the S-matrix in Eq. (\ref%
{s-matrix}), we find that two-qubit operations can be performed by the
collision process, where $U$ and relative group velocity $\upsilon _{\mathrm{%
r}}=\upsilon _{\text{\textrm{L}}}-\upsilon _{\text{\textrm{R}}}$\ are
connected to the evolution time by the relation%
\begin{equation}
t=\theta -\pi =2\cot ^{-1}\left( \frac{U}{\upsilon _{\mathrm{r}}}\right) .
\end{equation}%
Then we can implement entangling and swap gates for two flying qubits via
dynamic process. To demonstrate the result, we consider several typical
cases with $U=0$, $\infty $, and $\pm \left\vert \upsilon _{\text{\textrm{r}}%
}\right\vert $, which correspond to the operations of swap, standby, and
entanglement, respectively. The collision processes are illustrated
schematically in Fig. \ref{fig1}. The advantage of such a scheme is that the
temporal control is replaced by pre-engineered on-state interaction $U$.

In order to check the above conclusion, numerical simulation is performed.
We define the initial and target states as\
\begin{eqnarray}
\left\vert \Psi \left( 0\right) \right\rangle &=&\left\vert \mathrm{L}%
,p,\uparrow \right\rangle \left\vert \mathrm{R},q,\downarrow \right\rangle ,
\label{Psi_i} \\
\left\vert \Psi _{T}\right\rangle &=&e^{i\frac{\theta }{2}}(-i\sin \frac{%
\theta }{2}\left\vert \mathrm{L},q,\uparrow \right\rangle \left\vert \mathrm{%
R},p,\downarrow \right\rangle  \notag \\
&&+\cos \frac{\theta }{2}\left\vert \mathrm{L},q,\downarrow \right\rangle
\left\vert \mathrm{R},p,\uparrow \right\rangle ),  \label{Psi_t}
\end{eqnarray}%
where $\left\vert \Psi _{T}\right\rangle $\ possess the same relative
position but the exchanged momentum compared with the state $\left\vert \Psi
\left( 0\right) \right\rangle $\ as in Eq. (\ref{Psi_i}). On the other hand,
we consider the evolved state $\left\vert \Psi \left( t\right) \right\rangle
$\ for the initial state being $\left\vert \Psi \left( 0\right)
\right\rangle $\ driven by the Hamiltonian (\ref{F_Hub}), and caculate the
fidelity $\left\vert \left\langle \Psi _{T}\right\vert \Psi \left( t\right)
\rangle \right\vert $\ in Fig. \ref{fig2}. It is shown that when the state $%
\left\vert \Psi \left( 0\right) \right\rangle $\ evolves to the same
position with $\left\vert \Psi _{T}\right\rangle $, the fidelity $\left\vert
\left\langle \Psi _{T}\right\vert \Psi \left( t\right) \rangle \right\vert $%
\ is almost to $0$, which is in agreement with our previous theoretical
analysis.

\section{Multiple collision}

\label{sec_multiple}We apply our result to many-body system. Considering the
case that the initial state is consisted of many separable local particles
with the same group velocity, termed as many-particle wavepacket train
(MPWT), our result can be applicable if each collision time is exact known.
In this paper, we only demonstrate this by a simple example. We consider the
collision of two MPWTs with particle numbers $M$ and $N$ ($N\geqslant M$).
All the distances between two adjacent particles in two trains are
identical. The initial state is

\begin{figure}[tbp]
\includegraphics[ bb=34 278 573 773, width=0.47\textwidth, clip]{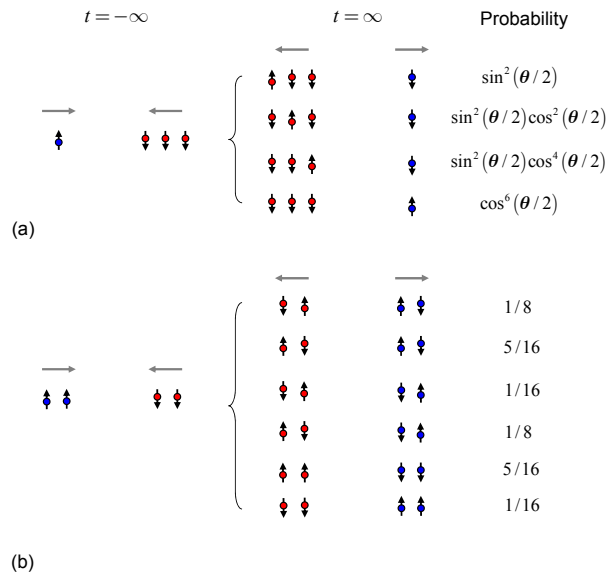}
\caption{(Color online) Schematic illustration of the collision between the
two MPWTs. (a) An incident single fermion comes from the left denoted as
blue spin and collides with $3$-fermion train, which comes from the right
denoted as red spins. It can be seen that the single fermion keep the
original momentum, but it entangles with the $3$-fermion train at the end of
the collision. The amplitudes of the four states are listed. It is shown
that the final state is direct product between the states of single fermion
and 3-fermion train when $\protect\theta =\protect\pi ,$ $\protect\theta =0$
with the corresponding parameter $U=\infty $, $0$, respectively. (b) The
collision between the two MPWTs come from the opposite direction with
particle number $N=2$. And the probability for the superposition of states
is listed with $\protect\theta =\protect\pi /2$.} \label{fig3}
\end{figure}

\begin{equation}
\prod\limits_{m=1}^{M}\left\vert \mathrm{L}_{m},p,\sigma _{m}\right\rangle
\prod\limits_{n=1}^{N}\left\vert \mathrm{R}_{n},q,\tau _{n}\right\rangle ,
\end{equation}%
where $\left\{ \mathrm{L}_{m}\right\} $ and $\left\{ \mathrm{R}_{n}\right\} $%
\ denote the sequences of particles, $\left\{ \sigma _{m}\right\} $ and $%
\left\{ \tau _{m}\right\} $\ denote the spin configurations in each trains.\
According to the above analysis, after collisions the final state has the
form of%
\begin{equation}
\prod\limits_{m=1}^{M}\left\vert \mathrm{L}_{m},q,\sigma _{m}^{\prime
}\right\rangle \prod\limits_{n=1}^{N}\left\vert \mathrm{R}_{n},p,\tau
_{n}^{\prime }\right\rangle ,
\end{equation}%
where the spin configurations $\left\{ \sigma _{m}^{\prime }\right\} $ and $%
\left\{ \tau _{m}^{\prime }\right\} $ are determined by the S-matrix, which
is the time-ordered product of all two-particle S-matrices. During the
collision process, the positions of particles in each train are always
spaced by equal intervals. This makes it easier to determine the times of
each collisions. Then the final state can be written as%
\begin{equation}
\prod\limits_{l=1}^{M}S_{l}\prod\limits_{n=1}^{N}\left\vert \mathrm{L}%
_{n},q,\tau _{n}\right\rangle \prod\limits_{m=1}^{M}\left\vert \mathrm{R}%
_{m},p,\sigma _{m}\right\rangle ,  \label{final}
\end{equation}%
where%
\begin{equation}
S_{l}=\prod\limits_{n=1}^{N}s_{l,N-n+1},
\end{equation}%
and%
\begin{equation}
s_{jk}=e^{-i\left( \theta -\pi \right) \left( \overrightarrow{\tau }%
_{j}\cdot \overrightarrow{\sigma }_{k}-1\right) /4},
\end{equation}%
where $\overrightarrow{\tau }_{j}\ $and $\overrightarrow{\sigma }_{k}$\ are
corresponding Pauli matrices. Applying the formula in Eq. (\ref{final}) to
the case with $M=1$, $\sigma _{1}=\uparrow $, $\tau _{n}=\downarrow $, $n\in %
\left[ 1,N\right] $, we obtain%
\begin{eqnarray}
&&\left\vert \mathrm{L}_{1},p,\uparrow \right\rangle
\prod\limits_{n=1}^{N}\left\vert \mathrm{R}_{n},q,\downarrow \right\rangle
\longmapsto  \notag \\
&&-i\sum_{j=1}^{N}e^{i\frac{\theta }{2}j}\sin \frac{\theta }{2}\cos ^{\left(
j-1\right) }\frac{\theta }{2}  \notag \\
&&\times \frac{1+\overrightarrow{\sigma }_{1}\cdot \overrightarrow{\tau }_{j}%
}{2}\prod\limits_{n}^{N}\left\vert \mathrm{L}_{n},q,\downarrow \right\rangle
\left\vert \mathrm{R}_{1},p,\uparrow \right\rangle  \notag \\
&&+e^{i\frac{\theta }{2}N}\cos ^{N}\frac{\theta }{2}\prod\limits_{n}^{N}%
\left\vert \mathrm{L}_{n},q,\downarrow \right\rangle \left\vert \mathrm{R}%
_{1},p,\uparrow \right\rangle .
\end{eqnarray}%
This conclusion is still true for the case with unequal-spaced $\left\{
\mathrm{R}_{n}\right\} $. For illustration, we sketch the case with $M=1$, $%
\sigma _{1}=\uparrow $, $N=3$, $\tau _{n}=\downarrow $, $n\in \left[ 1,3%
\right] $ in Fig. \ref{fig3}(a). One can see that the spin part of the final
state is the superposition of the combinations of the four spins.

Now we turn to investigate the entanglement between the single fermion and
the MPWT with particle number $N$. As is well known, the generation and
controllability of entanglement between distant quantum states have been at
the heart of quantum information processing. Such as the applications in the
emerging technologies of quantum computing and quantum cryptography, as well
as to realize quantum teleportation experimentally \cite{Nielson,Horodecki}.
Moreover, quantum entanglement is typically fragile to practical noise.
Every external manipulation inevitably induces noise in the system. This
suggests a scheme based on the above mentioned collision process for
generating the entanglement between a single fermion and the $N$-fermion
train without the need for the temporal control and measurement process. We
note that although the incident single fermion keep the original momentum,
it entangles with the $N$-fermion train after the collision, leading to a
deterioration of its purity. To measure the entanglement between the single
fermion and the $N$-fermion train, we calculate the reduced density matrix
of the single spin%
\begin{equation}
\rho _{\mathrm{R}}\left( \infty \right) =\left(
\begin{array}{cc}
\Lambda & 0 \\
0 & 1-\Lambda%
\end{array}%
\right) ,
\end{equation}%
where%
\begin{equation}
\Lambda =\cos ^{2N}\frac{\theta }{2}.
\end{equation}%
Thus the purity of the single fermion can be expressed as%
\begin{equation}
P(\infty )=\mathrm{Tr}\left( \rho _{\mathrm{R}}^{2}\right) =2\left( \Lambda -%
\frac{1}{2}\right) ^{2}+\frac{1}{2},
\end{equation}%
where Tr$\left( ...\right) $\ denotes the trace on the single fermion. For
the case of $\Lambda =0,$\ $1$, we have $P(\infty )=1$, which requires $%
\theta =\pi ,$\ and $\theta =0$, obtained from interaction parameter $%
U=\infty $, and $0$, respectively. It indicates that the single fermion
state and $N$-fermion train state are not entangled. In contrast, the purity
$P(\infty )=1/2$ at $\Lambda =1/2$ when
\begin{equation}
\theta =2\cos ^{-1}\left( 2^{-\frac{1}{2N}}\right) .  \label{theta}
\end{equation}%
It corresponds to a completely mixed state of the outgoing single fermion,
or maximal entanglement between the\ single fermion state and $N$-fermion
train. Together with Eq. (\ref{R}), we have%
\begin{equation}
U=\left( \upsilon _{\text{\textrm{R}}}-\upsilon _{\text{\textrm{L}}}\right)
\tan \left[ \cos ^{-1}\left( 2^{-\frac{1}{2N}}\right) \right] ,  \label{U_N}
\end{equation}%
which reduces to $U\approx (\upsilon _{\text{\textrm{R}}}-\upsilon _{\text{%
\textrm{L}}})\sqrt{\ln 2}N^{-1/2}$ for large $N$. This indicates that for
large $N$, one needs to take a small $U$\ of order $N^{-1/2}$\ to generate
the maximal entanglement between the\ single fermion state and $N$-fermion
train, or result in full decoherence of the single fermion.

In the case of two-train collision, the calculation can still be performed
in the similar way. However, it is hard to get analytical result for
arbitrary system. Here, we sketch the case with $M=2$, $\sigma _{1}=\sigma
_{2}=\uparrow $, $N=2$, $\tau _{1}=\tau _{2}=\downarrow $, in Fig. \ref{fig3}%
(b). The probability on each spin configuration is listed as illustration.

\section{Summary}

\label{sec_summary}In summary, we presented an analytical study for
two-fermion dynamics in Hubbard model. We find that the scattering matrix of
two-fermion collision is separable into two independent parts, operating on
spatial and spin degrees of freedom, respectively, when two incident
wavepackets have identical shapes. For two fermions with opposite spins, the
collision process can create a distant EPR pair due to the resonance between
the Hubbard interaction strength and the relative group velocity. The
advantage of this scheme is without the need of temporal control and
measurement process. Since it is now possible to simulate the Hubbard model
via cold fermionic atoms in optical lattice, these results can be realized
experimentally.

\acknowledgments We acknowledge the support of the National Basic Research
Program (973 Program) of China under Grant No. 2012CB921900 and the CNSF
(Grant No. 11374163). X. Z. Zhang is supported by PhD research startup
foundation of Tianjin Normal University under Grant No. 52XB1415.

\end{document}